\documentstyle[aps,prb]{revtex}

\def\SSC{Solid State Commun.\ }
\def\prep{Phys. Rep.\ }
\def\ZETF{Zh. Eksp. Teor. Fiz.\ }
\def\SPJ{Sov. Phys.-JETP\ }
\def\WRM{Waves Random Media\ }

\begin{document}

\draft

\title{Elastic Scattering as a Cause of Quantum Dephasing:\\
The Conductance of Two-Dimensional Imperfect Conductors}

\author{Yu.~V.~Tarasov}

\address{Institute for Radiophysics \& Electronics,
National Academy of Sciences of Ukraine,\\
12 Acad. Proskura St., Kharkov 61085, Ukraine}

\date{\today}

\maketitle

\begin{abstract}
A method is proposed for studying wave and particle transport in
disordered waveguide systems of dimension higher than unity by
means of exact one-dimensionalization of the dynamic equations in
the mode representation. As a particular case, the $T=0$
conductance of a two-dimensional quantum wire is calculated, which
exhibits ohmic behaviour, with length-dependent conductivity, at
any conductor length exceeding the electron quasi-classical mean
free path. The unconventional diffusive regime of charge transport
is found in the range of conductor lengths where the electrons are
commonly considered as localized. In quantum wires with more than
one conducting channel, each being identified with the extended
waveguide mode, the inter-mode scattering is proven to serve as a
phase-breaking mechanism that prevents interference localization
without real inelasticity of interaction.

\end{abstract}

\pacs{PACS numbers: 71.30.+h, 72.15.Rn, 73.50.-h}


\section{Introduction}
\label{intro}

Since the original formulation of the localization problem by
Anderson \cite{anders58}, the question of whether electronic
states in disordered systems are localized at any strength of
disorder or a mobility edge can be defined by a critical level of
disorder has become a central issue. In the former case this
yields an insulating-type behaviour of large samples, while in the
latter the metallic-type transport of conducting electrons is
allowed due to the existence of extended states.

It was subsequently ascertained that the answer to this question
depends substantially on the dimensionality of the disordered
system. For the case of one-dimensional (1D) conductors in the
limit of vanishing temperature one can prove in a mathematically
rigorous way (see Ref.~\onlinecite{LifGredPas} and references
therein) that the spectrum of the electrons, subject to an
arbitrarily small but finite random potential in infinitely long
samples, is purely discrete, i.e. all electron states are
necessarily localized irrespective of their energy. As a result,
the DC conductivity vanishes for such systems, whereas in finite
1D samples the conductance falls exponentially as the length of
the conductor grows.

In contrast to the 1D case, no mathematically rigorous theory of
localization exists for two- (2D) and tree-dimensional (3D) random
systems. Formulation of the problem in terms of the
renormalization group\cite{wegner76} (RG) led to the one-parameter
scaling hypothesis of localization\cite{abrah79} and appeared to
provide a considerable progress in studying systems of
dimensionality greater than unity. The main conviction gained from
this development was that all the electron states in both 1D and
2D disordered systems are localized at arbitrary small disorder.
Hence, the metal-insulator transition (MIT) is usually believed to
occur only in 3D systems, while (sufficiently large) 1D and 2D
systems are always Anderson insulators.\cite{myremark_1}

This opinion, established quite long ago, has recently been
challenged after unexpected experimental detection of MIT in
dilute 2D electron and hole systems.\cite{kravch9596} Originally
obtained on Si-MOSFETs, the results have entailed numerous
experimental findings of MIT in other dilute 2D systems,
\cite{popov97,coler97,simmon98,papad98,hanein98} leading to the
well-founded belief that the effect is of a rather general nature.
To elucidate the unconventional experimental facts, different
theoretical approaches were put forward including quite disputable
ones (see the discussion in Ref.~\onlinecite{aag99}). Among the
variety of options the greatest anticipations in explaining the
experiments are mostly associated with accounting for the Coulomb
interaction of carriers.\cite{shepelyan94,weinman97} Yet the
transport theories for correlated particles in the presence of
disorder still cannot claim for general acceptance because of the
substantial controversy in the interpretation of the role of
interaction within different parameter domains which correspond to
diffusive\cite{berkovits96,schmitteckert98} and
localized\cite{efros95,talamantes96,vojta98} regimes.

Traditional understanding of electron localization in 2D
disordered systems and numerous experimental facts which indicate
the metallic nature of charge transport at low temperatures have
brought about active progress in the line of study that is
identified with {\it quantum dephasing}. The main efforts in this
direction are focused on the detection of various {\it
phase-breaking mechanisms} responsible for the delocalization of
quasi-particles primarily localized by disorder.  However, in
spite of a large body of versions on offer the problem still
remains open.

Our intention in this contribution is to re-examine the
conventional one-particle approach without appealing to scattering
mechanisms other than those characteristic of quenched disorder.
Evidence will be given that even within this elementary model the
experimental findings of
Refs.~\onlinecite{kravch9596,popov97,coler97,simmon98,papad98,hanein98},
at first sight curious, are as a matter of fact quite natural.
Numerous attempts to interpret the results of
Ref.~\onlinecite{kravch9596} within the framework of a
single-particle approach were made, in particular, by improving
the scaling approach.\cite{dobrosavl97} In this study, however, a
fundamentally different strategy is chosen which is an alternative
to the RG analysis.\cite{ma76} We prefer to obtain the observables
directly, while conclusions (though indirect) about the
localization of electron states are made on the basis of the
results.

It is instructive to recall that working out, even without a
profound spectral analysis, practical asymptotic methods for
calculating the disorder-averaged many-particle characteristics
(conductivity, density-density correlator, etc.) turned out to be
more helpful for the establishment of a highly advanced theory of
1D random systems than the development of rigorous mathematical
foundations.\cite{berezinski73,abrikryzh78,kanercheb87,kanertar88}
The usefulness of such an approach can be attributed to the fact
that in the context of the above-mentioned essentially
perturbative methods one has managed to trace with the required
accuracy the effect of mutual interference of quantum waves
corresponding to multiply backscattered current carriers. In such
a way, physical results entirely consistent with the anticipations
based on mathematical predictions were eventually obtained. The
present research was primarily induced by long-standing discontent
associated with the lack of arguments of a comparable standard,
either in favour of localization or against it, as applied to 2D
systems of degenerate electrons subject to a static random
potential.

Commonly, the presence of inelastic scattering mechanisms is held
responsible as a main cause of preventing quantum interference
and, thus, Anderson localization.\cite{bu_im_lan83} Among these
are the electron-phonon and electron-electron interactions and
other conceivable methods of energy interchange between the
electron bath and the environment.\cite{mello99} These can lead to
the loss of phase (meaning energy) memory or, in other words,
phase coherence of electronic states. Note in this connection that
in the 1D case the demand of energy coherence admits a large
transfer of {\em momentum} for onefold scattering of degenerate
electrons in the backward direction. This leads inevitably to
considerable local breaking of {\it spatial} (instead of {\em
temporal}) phases of the wavefunctions. It was recognized that
such a large violation of spatial phases is quite helpful when
deriving a constructive theory of 1D quantum transport. On the
basis of such arguments, the selection of efficient subsets of
terms in perturbative expansion which are responsible for the
effective interference of electronic waves scattered iteratively
in a {\it backward} direction was suggested.\cite{berezinski73}
This interference finally results in the formation of true
localized states, even when one starts their perturbative
construction from plain-wave-like trial functions which belong to
a mathematical class different from that of localized functions.

An analogous scenario of the perturbative formation of localized
states in 2D systems cannot give the same result, since for
sufficiently isotropic scattering the spatial coherence of the
wavefunctions is already violated at distances of the order of the
quasi-classical mean free path even for weak energy scattering.
The coherence is maintained efficiently only within a small phase
volume, thus resulting in the relative smallness of the
interference corrections, usually known as weak localization
corrections\cite{altaronkhmel82,altarkhmellark82} for
non-one-dimensional systems.

Nevertheless, the analysis of the problem of multidimensional
localization with the use of a one-dimensional scenario turns out
to be quite productive. As we shall prove below the problem of
electron transport in 2D open system of waveguide type can be
reduced without any approximations to the set of one-dimensional
(though non-Hermitian) problems for the quantum waves propagating
in individual conducting channels. The channels will be identified
with extended waveguide modes. For the corresponding dynamic
equations that are one-dimensional, an opportunity arises for
making substantial use of {\em spatial} phases of the
wavefunctions instead of their {\em temporal} parts. This turns
out to be preferable from the technical point of view for solving
stationary problems of the electron, as well as classical wave,
transport. The reason for the usefulness of such a procedure lies
in the fact that in one-dimensional problems spatial averaging has
been shown to be highly advantageous, leading to the reduction of
the perturbative expansions of the physical observables to a
summable series.\cite{berezinski73,abrikryzh78}

Being exactly quantum in nature, the waveguide approach used here
is, to a certain degree, less obvious than semi-classical ones
usually applied in most localization theories. Thus, in its
context the clarity of the path-based interpretation is
substantially lost. At the same time, the benefit of our method is
that one-dimensional `channel' equations enable one to distinguish
unambiguously the coherent intra-mode scattering, which is easily
interpreted from the standpoint of 1D localization theory, and the
inter-mode scattering which corresponds, although not quite
directly, to isotropic scattering of semi-classical electrons. The
quantum states in different channels are specified by different
longitudinal momenta. This difference essentially suppresses
interference of primary and scattered electronic waves, if they
belong to different channels. As a result, the inter-mode
scattering turns out to be an intrinsic origin of the inability
for 2D electrons to be localized by weak static disorder.

The `one-dimensionalization' procedure suggested in this paper
gives an opportunity to highlight the role of spatial coherence in
the interference of electronic waves even in a single-particle
approximation.\cite{feynman65} The results obtained here by
conventional perturbative methods enable us to conclude about the
unrealizability of the strong (Anderson) localization in systems
whose spatial dimensionality is greater than unity, irrespective
of their size. It should be particularly emphasized that
decoherence appearing due to inter-mode scattering is unconnected
to genuine inelasticity in the interaction of electrons with
disorder. However, the difference in {\em longitudinal energy}
between the conducting channels could be treated as a source of
`hidden inelasticity', if one is accustomed to such an
interpretation.

The paper is organized as follows. In the next section, the
problem is formulated using linear response theory. In section
\ref{Uni_proc} we develop a method of exact one-dimensionalization
which is a central point of the paper. Then, in section
\ref{trial_green}, the trial Green functions supremely important
for the developed technique are analyzed with the aid of a
two-scale perturbation method. A spectral analysis of the electron
system is given in section \ref{Spectrum}. In the final two
sections we present asymptotic expressions for the conductance and
discuss the results. A pair of tedious but important calculations
is presented in two appendices.

\section{Statement of the problem}
\label{statement}

A common approach used in studies of random systems of various
dimensionality is to take a hyper-cube of a certain linear size and vary
the size while searching for the conductance scaling. Such an approach
seems to be natural when studying spectral properties of closed systems. At
the same time, it is not quite appropriate for solving transport problems
as applied to structures of waveguide type, in particular, quantum
conductors of arbitrary length.

In this paper, we examine the case of a 2D imperfect rectangular
sample of length $L$ in the $x$-direction and width $D$ in the
lateral direction $y$.  Degenerate non-interacting spinless
electrons are assumed to be confined between the hard-wall side
boundaries $y=\pm D/2$, whereas in the direction of current ($x$)
we suppose the system to be open at the strip ends $x=\pm L/2$.
The dimensionless conductance $g(L)$ (in units $e^2/\pi\hbar$) is
calculated directly from linear response theory,\cite{kubo57}
whence at zero temperature we have

\begin{equation}
g(L)=-\frac{4}{L^2}\int\!\!\!\!\int d\bbox{r}\,d\bbox{r}'
\frac{\partial G(\bbox{r},\bbox{r}')}{\partial x}\cdot
\frac{\partial G^*(\bbox{r},\bbox{r}')}{\partial x'} \ .
\label{Kubo_cond}
\end{equation}
Here the integration with respect to $\bbox{r}=(x,y)$ is performed over the
area occupied by the conductor

\begin{equation}
x\in(-L/2,L/2)\ ,\qquad y\in(-D/2,D/2) \ .
\label{cond_area}
\end{equation}
$G(\bbox{r},\bbox{r}')$ is the retarded Green function of the conducting
electrons. Within the isotropic Fermi liquid model this function is
governed by the equation

\begin{equation}
\left[ \Delta+k_F^2+i0-V(\bbox{r}) \right]G(\bbox{r},\bbox{r}')=
\delta(\bbox{r}-\bbox{r}') \ .
\label{StartEq}
\end{equation}
We adopt the system of units with $\hbar=2m=1$ ($m$ is the electron
effective mass), so that $\Delta$ is a two-dimensional Laplace operator,
$k_F$ is the Fermi wavenumber of the electrons, $V(\bbox{r})$ is the `bulk'
static random potential.

The potential $V(\bbox{r})$ in equation (\ref{StartEq}) will be
regarded as a short-range one and not necessarily isotropic. The
term `short-range' implies the characteristic spatial interval
over which the potential is substantially varied to be small
compared with the `macroscopic' lengths of the problem, namely the
electron mean free path and the conductor length. Being considered
as a statistical variable, the potential $V(\bbox{r})$ will be
specified by a zero mean value and the binary correlation function

\begin{equation}
\left< V(\bbox{r})V(\bbox{r}') \right>={\cal Q}W(\bbox{r}-\bbox{r}') \ .
\label{VrVr}
\end{equation}
Here the angular brackets denote ensemble averaging and
$W(\bbox{r})$ is some function normalized to unity. The explicit
form of this function is not so important. In many cases
$W(\bbox{r})$ is approximated by the delta-function,
$W(\bbox{r})=\delta(\bbox{r})$. However, the method developed in
this contribution permits one to consider not only isotropic and
not necessarily local scattering events. In addition, the choice
of $W(\bbox{r})$ in the form of a delta-function, apart from
restricting the physical applicability of the model, is not quite
convenient from the technical viewpoint. Thus, for example, when
calculating the corrections to the mode spectrum, equations
(\ref{T-renorm}), a formal problem can arise of the divergence of
the evanescent mode contribution. It is certainly absent provided
the potential $W(\bbox{r})$ is not exactly point-supported ---
this problem is familiar in quantum
mechanics.\cite{BazZeldPerel,LandauLif} To get rid of the
divergences in the problems of dimension more than one it is
merely sufficient to choose the function $W(\bbox{r})$ to be less
singular than $\delta(\bbox{r})$.\cite{LifGredPas} Therefore, in
order to focus on the main questions, in place of the correlation
equality (\ref{VrVr}), we shall use the expression below

\begin{equation}
\left< V(\bbox{r})V(\bbox{r}') \right>={\cal Q}W(x-x')\delta(y-y') \ .
\label{VrVr_2}
\end{equation}
Additionally, this form of equation (\ref{VrVr}) allows us to
consider anisotropic scattering. Similar to $W(\bbox{r})$ from
(\ref{VrVr}), the function $W(x)$ in (\ref{VrVr_2}) is normalized
to unity.

\section{The one-dimensionalization procedure}
\label{Uni_proc}

\subsection{The general scheme}
\label{UNI-GEN}

It seems intuitively natural for an open system with the
prescribed direction of quasi-particle transport to be considered
as to some extent a one-dimensional object. However, the
realizability of `one-dimensionalization' at the level of dynamic
equations, which would be quite important from a mathematical
perspective, is not {\it a priori} apparent. Here the term
`one-dimensionalization' means reduction of the two-dimensional
differential equation (\ref{StartEq}) to a set of one-dimensional
equations.  Although a system of waveguide type can often be
regarded as a collection of one-dimensional quantum channels, the
latter are not independent in general. Normally, they are strongly
coupled with each other through static or dynamic inhomogeneities
present in the problem.

Nevertheless, in what follows we intend to show that just the
waveguide nature of a system under consideration enables the
mathematical description of the transport problem for a 2D region
(\ref{cond_area}) to be reduced to a set of independent strictly
one-dimensional, although non-Hermitian, problems posed on the
interval $x\in(-L/2,L/2)$, regardless of the strength of the
disorder. To perform the reduction one should merely pass to the
mode representation, i.e.  Fourier transform in the transverse
coordinate $y$, using some complete set \{$\phi(y)$\} of
eigenfunctions of the transverse free-electron Hamiltonian, namely
the Laplace operator in Eq.~(\ref{StartEq}). The conductance
(\ref{Kubo_cond}) acquires the form

\begin{equation}
g(L)=
- \frac{4}{L^2} \int\!\!\!\!\int_L dx dx'\sum_{n,n'=1}^{N_c}
\frac{\partial G_{nn'}(x,x')}{\partial x}
\frac{\partial G_{nn'}^*(x,x')}{\partial x'} \ ,
\label{Cond-mode}
\end{equation}
where $N_c=[k_FD/\pi]$ is the number of conducting channels or, in
other words, extended waveguide modes. Equation (\ref{StartEq}) is
then transformed into a set of coupled ordinary differential
equations for the Fourier components $G_{nn'}(x,x')$,

\begin{equation}
\bigg[ \frac{\partial^2}{\partial x^2}+k_n^2+i0
- V_n(x)\bigg]G_{nn'}(x,x')
-\sum_{m=1\atop(m\neq n)}^{\infty} U_{nm}(x)G_{mn'}(x,x')
=\delta_{nn'}\delta(x-x') \ .
\label{ModeEq}
\end{equation}
Here $k_n^2=k_F^2-(n\pi/D)^2$ is the longitudinal energy of the $n$th mode,
$U_{nm}(x)$ is the inter-mode matrix element of the potential $V(\bbox{r})$,

\begin{equation}
U_{nm}(x)=\int_D dy\, \varphi_n(y)V(\bbox{r})\varphi_m(y) \ .
\label{Unm}
\end{equation}

Note the difference between the summation limits in
Eqs.~(\ref{Cond-mode}) and (\ref{ModeEq}). Restriction of the
summation in (\ref{Cond-mode}) by the number of conducting
channels implies, strictly speaking, weakness of the electron
scattering, to be specified in Sec.~\ref{WEAK-APPROX}. Under the
same assumption formula (\ref{Kubo_cond}) itself is valid, where
the products of the Green functions of the same kind (both
retarded and advanced) have already been omitted. In equation
(\ref{ModeEq}), the summation is naturally performed over a
complete set of waveguide modes.

Particular attention should be drawn to the fact that the term
containing the diagonal (intra-mode) matrix element
$U_{nn}(x)\equiv V_n(x)$ is initially detached from the sum of
Eq.~(\ref{ModeEq}), so that the matrix $\|U_{nm}\|$ hereafter is
held off-diagonal in the discrete mode variable. This little
technical trick enables one to reduce the problem of finding the
overall of the functions $G_{nn'}(x,x')$ to the solution of a
subset of purely one-dimensional closed equations for the
mode-diagonal functions $G_{nn}(x,x')$ only. To this end we first
introduce the auxiliary trial Green functions $G_n^{(V)}(x,x')$
($n\in\aleph$), each obeying the equation

\begin{equation}
\left[\frac{\partial^2}{\partial x^2}+k_n^2+i0- V_n(x)\right]
G_n^{(V)}(x,x')=\delta(x-x') \ ,
\label{trial_Gn}
\end{equation}
and Sommerfeld's radiation
conditions\cite{BassFuks79,Vladimirov67} at the strip ends $x=\pm
L/2$. These conditions seem to be natural to impose on an {\em
open} system. For the case of the 1D equation (\ref{trial_Gn})
they acquire the form\cite{Klyatskin86}

\begin{equation}
\left.\left(\frac{\partial}{\partial x}\mp ik_n\right)
G_n^{(V)}(x,x')\right|_{x=\pm L/2}=0  \ .
\label{1Dradcond}
\end{equation}
It implies that the field of the $n$th mode radiated by the point
source placed at $x'\in(-L/2,L/2)$ reaches the corresponding
($\pm$) end of the interval and then propagates unscattered with
the conserved momentum $k_n$ beyond the ends of the conductor.
Possible scattering in the leads attached to the strip from the
left and right should be taken into account separately.

Although a solution of the stochastic problem (\ref{trial_Gn}) and
(\ref{1Dradcond}) cannot be obtained in quadratures, in the case
of weak scattering specified below by inequalities
(\ref{weakscat}) the main features of the solution can be
extracted by due consideration with any desired accuracy in the
{\em intra-mode} potential $V_n(x)$. The necessary analysis is
presented in the next section. As for now, we merely consider all
the functions $G_n^{(V)}(x,x')$ as {\em a priori} known ones whose
properties are specified solely by the elastic intra-mode
scattering. With these functions chosen as an initial
approximation for the exact mode functions $G_{nn}$, only the
inter-mode scattering associated with the off-diagonal matrix
$U_{nm}(x)$ will then be taken as a perturbation. To implement
this intent, we turn from equation (\ref{ModeEq}) to the
consequent integral equation,

\begin{equation}
G_{nn'}(x,x')=G_n^{(V)}(x,x')\delta_{nn'}
+\sum_{m=1}^{\infty}\int_L dx_1\,
{\sf R}_{nm}(x,x_1)G_{mn'}(x_1,x') \ .
\label{Mode_inteq}
\end{equation}
Here the kernel
\begin{equation}
{\sf R}_{nm}(x,x_1)=G_n^{(V)}(x,x_1)U_{nm}(x_1)
\label{kern_ R}
\end{equation}
already contains only those harmonics of the potential
$V(\bbox{r})$ which are responsible for the inter-mode scattering.
A thorough study of system (\ref{Mode_inteq}) leads to a
conjecture that all non-diagonal mode elements $G_{mn}$ ($m\neq
n$) can be expressed only via the diagonal element $G_{nn}$ by
means of some linear operator $\hat{\sf K}$,

\begin{equation}
G_{mn}(x,x')=\int_Ldx_1\,{\sf K}_{mn}(x,x_1)G_{nn}(x_1,x') \ .
\label{G_mn-sol}
\end{equation}
To specify this operator, one should separate the term with the
diagonal (in the mode variable) Green function on the right-hand
side of equation (\ref{Mode_inteq}) and substitute all
non-diagonal Green functions in the form of equation
(\ref{G_mn-sol}). Then, after renaming the mode variables, we
arrive at the following equation for the matrix elements of the
operator $\hat{\sf K}$:

\begin{equation}
{\sf K}_{mn}(x,x')={\sf R}_{mn}(x,x')
+\!\!\sum_{k=1\atop (k\neq n)}^{\infty}
\int_Ldx_1\,{\sf R}_{mk}(x,x_1){\sf K}_{kn}(x_1,x') \ .
\label{K_mn}
\end{equation}

Equation (\ref{K_mn}) belongs to a class of multi-channel
Lippmann-Schwinger equations that are known to be extremely
singular, in contrast to their single-channel
counterparts.\cite{Taylor72} Nevertheless, by choosing the trial
Green function $G_n^{(V)}$ as a zero approximation and perturbing
it only by the inter-mode potentials, one manages to avoid the
above mentioned singularity.  Note that mode indices $m$ and $k$
in Eq.~(\ref{K_mn}) take all the positive integer values except
for the value $n$. This urges one to interpret the functions
appearing in (\ref{K_mn}) as matrix elements of operators acting
in two-dimensional mixed coordinate-mode space ($x,m$) which does
not include the mode $n$ (the notation ${\sf{\overline M}_n}$ will
be used for that space). The presence in Eq.~(\ref{K_mn}) of the
right-hand index $n$, which does not belong to ${\sf{\overline
M}_n}$, can be ensured by introducing the projection operator
${\bf P}_n$ that will make the mode index of any operator standing
next to it (both from the left or right) equal to $n$. With this
convention accepted, it follows from equation (\ref{K_mn}) that
the operator $\hat{\sf K}$ implementing relation (\ref{G_mn-sol})
has the form

\begin{equation}
\hat{\sf K}=\left(\openone-\hat{\sf R}\right)^{-1}\hat{\sf R}{\bf P}_n \ .
\label{HAT_K}
\end{equation}
Here $\hat{\sf R}$ is a 2D operator acting on ${\sf{\overline
M}_n}$ and is specified by the matrix elements (\ref{kern_ R}).

As for the mathematical correctness of the operator representation
(\ref{HAT_K}), it depends on the existence of the inverse operator
$\left(\openone-\hat{\sf R}\right)^{-1}$. This point is discussed
in Appendix \ref{K_exist}, where we provide a proof that
detachment of the intra-mode potential $V_n(x)$ in
Eq.~(\ref{ModeEq}) prevents the possible singularity.

Thus, equations (\ref{G_mn-sol}) and (\ref{HAT_K}) reduce the
problem of finding the complete Green function
$G(\bbox{r},\bbox{r}')$ within the 2D region (\ref{cond_area}) to
calculation of its diagonal mode components only. In order to do
that it is necessary to put $n'=n$ in Eq.~(\ref{ModeEq}) and
substitute all non-diagonal components $G_{mn}$ from equation
(\ref{G_mn-sol}). As a result, the closed equation  for the
diagonal component $G_{nn}$ is deduced,

\begin{equation}
\left[\frac{\partial^2}{\partial x^2}+k_n^2
+i0-V_n(x)-\hat{\cal T}_n\right]
G_{nn}(x,x')=\delta(x-x') \ .
\label{1Deq}
\end{equation}
In equation (\ref{1Deq}), in addition to the {\em local}
intra-mode potential $V_n(x)$, the {\em operator} potential
${\hat{\cal T}}_n$ has arisen,

\begin{equation}
{\hat{\cal T}}_n={\bf P}_n\hat{\cal U}\left(\openone-
\hat{\sf R}\right)^{-1}\hat{\sf R}{\bf P}_n =
{\bf P}_n \hat{\cal U}\left(\openone-
\hat{\sf R}\right)^{-1}{\bf P}_n \ .
\label{Tn}
\end{equation}
Here $\hat{\cal U}$ is the inter-mode operator specified on
${\sf{\overline M}_n}$ by the matrix elements

\begin{equation}
\biglb<x,l\bigrb|\hat{\cal U}\biglb|x',m\bigrb>=U_{lm}(x)\delta(x-x') \ .
\label{calU}
\end{equation}

The potential ${\hat{\cal T}}_n$ has quite an interesting
interpretation. In the right-hand side of equation (\ref{Tn})
there is a conventional $T$-matrix\cite{Taylor72,Newton68}
enveloped by the projective operators, one of which removes the
excitation from mode $n$ and the other restores it back to the
same mode after the appropriate scattering events within the
subspace ${\sf{\overline M}_n}$. Hence, the potential ${\hat{\cal
T}}_n$, although corresponding to effectively intra-mode
scattering, actually includes all inter-mode scattering events
undergone by the excitation while passing over `closed paths' in
the mode space. The intra- and inter-mode scattering mechanisms
turn out to be attributed to different potentials in equation
(\ref{1Deq}), facilitating significantly the subsequent
interpretation of the results. In what follows the potentials
$V_n(x)$ and ${\hat{\cal T}}_n$ will be referred to as those
responsible for {\em direct} intra-mode and inter-mode scattering,
respectively.

Finally, in this subsection we express the conductance
(\ref{Cond-mode}) through the diagonal elements of the mode Green
matrix. After rearranging the terms in Eq.~(\ref{Cond-mode}) and
using relation (\ref{G_mn-sol}) we obtain

\begin{eqnarray}
g(L)=-\frac{4}{L^2}\sum_{n=1}^{N_c}
&&\int\!\!\!\!\int_L dx\,dx'\Bigg[
\frac{\partial G_{nn}(x,x')}{\partial x}
\cdot\frac{\partial G_{nn}^*(x,x')}{\partial x'}
\nonumber \\
+\sum_{m=1\atop(m\neq n)}^{N_c}
&&\int\!\!\!\!\int_L dx_1dx_2\,
\frac{\partial {\sf K}_{mn}(x,x_1)}{\partial x}G_{nn}(x_1,x')
{\sf K}_{mn}^*(x,x_2)\frac{\partial G_{nn}^*(x_2,x')}{\partial x'}\Bigg]\ .
\label{Cond_2}
\end{eqnarray}
Expression (\ref{Cond_2}) jointly with equation (\ref{1Deq})
completes, in principle, the `one-dimensionalization' procedure
introduced at the beginning of this section. In this form the
problem under consideration is convenient for a  numerical
treatment at any disorder strength, because the solution of the 2D
problem governed by equation (\ref{StartEq}) is reduced to a {\em
finite set} of purely 1D problems (\ref{trial_Gn}) and
(\ref{1Deq}). At the same time, assuming weak electron scattering
(in the semi-classical sense), we manage to proceed with our
analytical consideration and obtain the results.

\subsection{The weak scattering approximation}
\label{WEAK-APPROX}

It is natural to specify the intensity of electron scattering in
terms of characteristic spatial scales inherent to the problem.
Henceforth we recognize the scattering as weak provided the
following inequalities hold:

\begin{equation}
k_F^{-1},r_c\ll\ell \ .
\label{weakscat}
\end{equation}
Here $r_c$ is the correlation radius of the potential
$V(\bbox{r})$, $\ell=2k_F/{\cal Q}$ is the {\it semiclassical}
mean free path of electrons evaluated within the model of a
$\delta$-correlated 2D random potential, i.e.
$W(\bbox{r})=\delta(\bbox{r})$ in Eq.~(\ref{VrVr}).

Estimation of the norm of the operator $\hat{\sf R}$ specified on
${\sf{\overline M}_n}$ by the matrix elements (\ref{kern_ R}) results in

\begin{equation}
\|\hat{\sf R}\|^2\sim \frac{D/L}{k_F\ell} \ .
\label{normR}
\end{equation}
Under conditions (\ref{weakscat}), this enables us to find an
expansion to lowest order in the impurity potential of the
operator $\hat{\sf K}$, Eq.~(\ref{HAT_K}), so that it becomes
approximately equal to $\hat{\sf R}$ almost regardless of the
conductor aspect ratio. The exact operator ${\hat{\cal T}}_n$ from
(\ref{Tn}) can, in turn, be replaced by the approximate value

\begin{equation}
{\hat{\cal T}}_n\approx {\bf P}_n\hat{\cal U}{\hat
G}^{(V)}\hat{\cal U}{\bf P}_n
\label{T_approx}
\end{equation}
where the operator ${\hat G}^{(V)}$ is specified on ${\sf{\overline M}_n}$
by the matrix elements
$$
\left<x,k\right|{\hat G}^{(V)}\left|x',m\right>=
\delta_{km}G_m^{(V)}(x,x') \ .
$$

Besides the reduction of the $T$-operator (\ref{Tn}) to truncated
form (\ref{T_approx}), a substitution of the approximate matrix
elements of the operator $\hat{\sf K}$ brings the second term of
conductance (\ref{Cond_2}) to the form

\begin{equation}
\int\!\!\!\!\int_L dx_1dx_2\, U_{mn}(x_1)U_{mn}(x_2)
\frac{\partial G_m^{(V)}(x,x_1)}{\partial x}{G_m^{(V)}}^*(x,x_2)
G_{nn}(x_1,x')\frac{\partial G_{nn}^*(x_2,x')}{\partial x'} \ ,
\label{aux_g2}
\end{equation}
which is convenient for performing the ensemble averaging. It will be
shown below that all Green functions in (\ref{aux_g2}), and not only the
trial ones, may be thought of as independent of the inter-mode
potentials $U_{mn}$. Yet correlation of those potentials with the
intra-mode one, $V_n(x)$, governing the trial Green functions, can be
disregarded in view of Eq.~(\ref{inter-intra}). As a result, after
averaging conductance (\ref{Cond_2}) with the use of (\ref{aux_g2})
and (\ref{VrVr_2}), the expression, which will be subject to a further
analysis, takes the form

\begin{eqnarray}
\big<g(L)\big>=-&&\frac{4}{L^2}\!\sum_{n=1}^{N_c}
\!\int\!\!\!\!\int_L\!\!dx\,dx'\!\left[
\Big<\frac{\partial G_{nn}(x,x')}{\partial x}
\frac{\partial G_{nn}^*(x,x')}{\partial x'}\Big> \right.
\nonumber \\
+&&
\frac{{\cal Q}}{D}\sum_{m=1\atop(m\neq n)}^{N_c}
\int\!\!\!\!\int_Ldx_1\,dx_2 \left.W(x_1-x_2)
\Big<{G_m^{(V)}}^*(x,x_2)\frac{\partial}{\partial x}G_m^{(V)}(x,x_1)\Big>
\Big<G_{nn}(x_1,x')\frac{\partial}{\partial x'}
{G_{nn}}^*(x_2,x')\Big>\right] \ .
\label{G(L)_AV}
\end{eqnarray}

\section{Analysis of the trial Green functions}
\label{trial_green}

The trial Green functions $G_m^{(V)}(x,x')$ enter the potential
${\hat{\cal T}}_n$, Eq.~(\ref{T_approx}), thus determining the
exact mode functions $G_{nn}(x,x')$, and the second term of the
conductance (\ref{G(L)_AV}). Although these functions appear as
subsidiary mathematical objects, they are of great concern for the
problem and therefore deserve particular consideration. The study
of the trial functions is also instructive, since it may provide
useful insights into the analysis of some misinterpretation
regarding 2D localization which existed until recently in the
literature.

Note first that the perturbative solution of equation
(\ref{trial_Gn}) depends substantially on whether the
corresponding unperturbed waveguide mode is either extended or
evanescent. The Green functions of evanescent modes with $n>N_c$
are localized even without any perturbation. To find them in the
limit of weak scattering, it is sufficient to restrict oneself to
zero-order perturbation in the potential $V_n(x)$,

\begin{equation}
G_n^{(V)}(x,x')= -\frac{1}{2|k_n|}\exp\biglb(-|k_n||x-x'|\bigrb) \ ,
\hspace{1.5cm} n>N_c \ .
\label{G_evan}
\end{equation}

The problem is much more involved for the extended modes, $n<
N_c$. Inasmuch as the function $G_n^{(V)}(x,x')$ is defined as a
solution of the strictly one-dimensional problem (\ref{trial_Gn})
and (\ref{1Dradcond}), to find it correctly in the context of
localization theory the plain-wave-based zero approximation is not
quite appropriate. This stems from the fact that in such an
approximation it is rather difficult to account for the
interference of multiply backscattered waves. Instead we apply the
two-scale perturbation method analogous to that used in the theory
of non-linear oscillations\cite{BogolMitr74}. This method showed
itself well for solving the problem of charge transport in
extremely narrow, namely single-mode, surface-corrugated
conducting strips.\cite{MakTar98} Below an outline of the method
is given together with some essential results. The details of
their derivation are deferred to Appendix~\ref{MOMENTS-GN}.

The method used in Ref.~\onlinecite{MakTar98} is based on the
representation of the 1D Green function, which is the solution of
the boundary-value problem (\ref{trial_Gn}) and (\ref{1Dradcond}),
via the solutions of the appropriate Cauchy problems,

\begin{equation}
G_n^{(V)}(x,x')=\bbox{\cal W}_n^{-1}
\big[ \psi_+(x|n)\psi_-(x'|n)\Theta(x-x') +
\psi_+(x'|n)\psi_-(x|n)\Theta(x'-x) \big] \ ,
\label{Green-Cochi}
\end{equation}
Here the functions $\psi_{\pm}(x|n)$ are linearly independent
solutions of the homogeneous equation (\ref{trial_Gn}) with the
radiation conditions analogous to (\ref{1Dradcond}) satisfied at
the strip ends $x=\pm L/2$, according to the `sign' index. The
Wronskian of those functions is $\bbox{\cal W}_n$, $\Theta(x)$ is
Heaviside's step function. This reformulation of a boundary-value
problem in terms of an initial-value problem will allow one to
perform averaging over the disorder later on.\cite{Klyatskin86}

It is advantageous to represent the functions $\psi_{\pm}(x|n)$ as
superpositions of modulated harmonic waves propagating in opposite
directions along the $x$-axis,

\begin{equation}
\psi_{\pm}(x|n) = \pi_{\pm}(x|n)\exp(\pm ik_n x) -
i\gamma_{\pm}(x|n)\exp(\mp ik_n x) \ .
\label{psi-pm}
\end{equation}
Within the framework of the weak scattering approximation
(\ref{weakscat}), the amplitudes $\pi_{\pm}(x|n)$ and
$\gamma_{\pm}(x|n)$ vary slowly as compared with the `fast'
exponentials $\exp(\pm ik_n x)$, so that the radiation conditions
for $\psi_{\pm}(x|n)$ are reformulated as the `initial' conditions
for the smooth amplitudes as follows:

\begin{equation}
\pi_{\pm}(\pm L/2|n) = 1 \ , \qquad \qquad \gamma_{\pm}(\pm L/2|n) = 0 \ .
\label{In_cond}
\end{equation}
In view of the smoothness of $\pi_{\pm}$ and $\gamma_{\pm}$,
differential equations for them can be derived by means of
averaging the equations for $\psi_{\pm}(x|n)$ over an
arbitrary-valued spatial interval intermediate between the
`microscopic' lengths $k_n^{-1}$ and $r_c$ on the one hand, and
the `macroscopic' lengths on the other hand. Among the latter
lengths are the scattering length, to be specified in the course
of the solution, and the sample length $L$. For weak scattering,
the equations for $\pi_{\pm}$ and $\gamma_{\pm}$ are reduced to
the coupled first-order ones,

\begin{equation}
\begin{array}{ccl}
&&\pi'_{\pm}(x|n)\pm i\eta_n(x)\pi_{\pm}(x|n)
\pm\zeta_{n\pm}^*(x)\gamma_{\pm}(x|n) =0\ , \\[6pt]
&&\gamma'_{\pm}(x|n)\mp i\eta_n(x)\gamma_{\pm}(x|n)
\pm\zeta_{n\pm}(x)\pi_{\pm}(x|n) =0\ .
\end{array}
\label{PiGamma_eq}
\end{equation}
Here the variable coefficients $\eta_n(x)$ and $\zeta_{n\pm}(x)$ are
random fields associated with the intra-mode potential $V_n(x)$ in the
following way:

\begin{equation}
\eta_n(x)=\frac{1}{2k_n}\int\limits_{x-l}^{x+l}\frac{dt}{2l}V_n(t) \ ,
\hspace{1.5cm}
\zeta_{n\pm}(x)=\frac{1}{2k_n}\int\limits_{x-l}^{x+l}\frac{dt}{2l}
{\rm e}^{\mp 2ik_nt}V_n(t) \ .
\label{EtaZeta}
\end{equation}
For the intermediate property of the averaging interval $2l$, the
fields $\eta_n(x)$ and $\zeta_{n\pm}(x)$ are, in fact, nothing but
the `narrow' sets of spatial harmonics of the potential $V_n(x)$
with the momenta close to zero and $\pm 2k_n$, respectively. The
real field $\eta_n(x)$ is responsible for the `forward'
scattering, while the complex one $\zeta_{n\pm}(x)$ for the
`backward' scattering of the $n$th waveguide mode.

Under the assumption of weak scattering, only binary correlators
of the random potentials govern the majority of statistical
characteristics of physical quantities. It was shown in
Ref.~\onlinecite{MakTar98} that only the two correlation
functions, $\left<\eta_n(x)\eta_n(x')\right>$ and
$\left<\zeta_{n\pm}(x)\zeta_{n\pm}^*(x')\right>$, of modified
random fields (\ref{EtaZeta}) may be thought of as non-vanishing.
Calculation with the use of model (\ref{VrVr_2}) readily gives

\begin{mathletters}
\label{Eta_Zeta}
\begin{eqnarray}
\left<\eta_n(x)\eta_n(x')\right>&=&
\frac{1}{L_f^{(V)}(n)}F_l(x-x') \ ,
\label{EtaEta}\\
\left<\zeta_{n\pm}(x)\zeta_{n\pm}^*(x')\right>&=&
\frac{1}{L_b^{(V)}(n)}F_l(x-x') \ .
\label{ZetaZeta}
\end{eqnarray}
\end{mathletters}
Here

\begin{equation}
L_f^{(V)}(n)=\frac{2D}{3{\cal Q}}(2k_n)^{2}
\qquad\text{and}\qquad
L_b^{(V)}(n)=\frac{2D}{3{\cal Q}}\frac{(2k_n)^{2}}{\widetilde{W}(2k_n)}
\label{Lf(n)Lb(n)}
\end{equation}
are the forward and backward mode scattering lengths, respectively;
$\widetilde{W}(q)$ is the Fourier transform of $W(x)$ from
Eq.~(\ref{VrVr_2}).  The function

\begin{equation}
F_l(x)=\int\limits_{-\infty}^{\infty}\frac{dq}{2\pi}{\text e}^{iqx}
\frac{\sin^2(ql)}{(ql)^2}=
\frac{1}{2l}\left(1-\frac{|x|}{2l}\right)\Theta(2l-|x|)
\label{F_l}
\end{equation}
in Eqs.~(\ref{Eta_Zeta}) is sharp at the scale of `macroscopic'
lengths, so it can be regarded as the $\delta$-function in the
`distributional' sense, $F_l(x)\to\delta(x)$.

Equations (\ref{PiGamma_eq}) and correlation relations
(\ref{Eta_Zeta}) enable one to obtain the entire statistical information
concerning the trial Green function for any extended mode $n$. In
Appendix~\ref{MOMENTS-GN} derivation of all plain moments of that
function is presented with the following result:

\begin{eqnarray}
&&\Big<\left[G_n^{(V)}(x,x')\right]^\mu\Big>=
\left(\frac{-i}{2k_n}\right)^\mu
\exp\left[i\mu k_n|x-x'|-\frac{\mu}{2}
\left(\frac{\mu}{L_f^{(V)}(n)}+
\frac{1}{L_b^{(V)}(n)}\right)|x-x'|\right] \ ,
\nonumber \\
&&\mu\in\aleph \ .
\label{moments}
\end{eqnarray}
It is noteworthy that from Eqs.~(\ref{Lf(n)Lb(n)}) we have the
following estimate:

\begin{equation}
L_{f,b}^{(V)}(n)\sim N_c\ell\cos^2\vartheta_n \ ,
\label{Lfb_estim}
\end{equation}
where $\vartheta_n$ is a `sliding angle' of the $n$th mode with respect to
the $x$-axis, $|\sin\vartheta_n|=n\pi/k_FD$. The value of (\ref{Lfb_estim})
is coincident, in order of magnitude, with the {\em localization length}
widely believed to be characteristic for multi-mode quasi-one-dimensional
(Q1D) waveguide systems (see, e.g.,
Refs.\onlinecite{Dorokhov84,MelPerKum88,TamAndo,YangRammer}). However, in
the present theory lengths (\ref{Lf(n)Lb(n)}) are nothing but the
extinction lengths of the {\em auxiliary} excitations that are not
subjected to inter-mode scattering.

Besides plain moments (\ref{moments}), important characteristics of
the random function $G_n^{(V)}(x,x')$ are the mixed moments
$\Big<\left[G_n^{(V)}(x,x')\right]^\mu
\left[{G_n^{(V)}}^*(x,x')\right]^\nu\Big>$. At $\mu=\nu$ all of
them are smooth (not oscillatory) functions of the argument $|x-x'|$ whose
spatial decrease is determined by the {\em one-dimensional localization
length} $\xi_n=4L_b^{(V)}(n)$.\cite{LifGredPas} The second term in
(\ref{G(L)_AV}) contains one of the simplest correlators of this type,
the `density-current' correlator $\big<G^*\nabla G\big>$. It was already
studied in applications to the problem of classical wave
transport.\cite{FreylTaras91} We do not present here the exact expression
for this correlator since only the fact of its exponential decrease at the
localization length $\xi_n$ is of significance for our analysis,

\begin{equation}
\Big<{G_m^{(V)}}^*(x,x')\frac{\partial}{\partial x}G_m^{(V)}(x,x')\Big>
\propto\exp\left(-\frac{|x-x'|}{\xi_n}\right) \ .
\label{corr_dens-curr}
\end{equation}

\section{The mode states spectrum: Perturbative treatment}
\label{Spectrum}

For calculation of the diagonal Green function $G_{nn}(x,x')$ in
equation (\ref{1Deq}) by means of the perturbation technique, it
would make a good sense to reconstruct the operator potential
${\hat{\cal T}}_n$ so that the mean value equals zero.  To that
end one has to define the result of the action of the operator
$\langle {\hat{\cal T}}_n\rangle$ on the function $G_{nn}$. For
$n>N_c$ (evanescent modes) the function $G_{nn}$ can be left in
its unperturbed form (\ref{G_evan}) due to the weakness of
scattering. For $n< N_c$, with the operator nature of the
potential $\langle {\hat{\cal T}}_n\rangle$ and reduced form
(\ref{T_approx}) of the $T$-operator taken into account, the
following integral has to be calculated:

\begin{equation}
\langle {\hat{\cal T}}_n\rangle G_{nn}(x,x')=
\int_L dx_1 {\sf T}_n(x,x_1)G_{nn}(x_1,x') \ .
\label{hat_T-Gnn}
\end{equation}
Here the kernel ${\sf T}_n(x,x_1)$ is given by

\begin{equation}
{\sf T}_n(x,x_1)=\sum_{m=1\atop(m\neq n)}^{\infty}
\left<U_{nm}(x)G_m^{(V)}(x,x_1)U_{mn}(x_1)\right> \ .
\label{krn_<T>}
\end{equation}
On performing the ensemble averaging in Eq.~(\ref{krn_<T>}), it is
justifiable to neglect the correlation between the inter-mode
potentials $U_{nm}(x)$ and the intra-mode ones, $V_m(x)$. Within
the model of point-like scatterers such correlation is entirely
absent. Yet even in the case of disorder specified by correlation
function (\ref{VrVr_2}) it is not difficult, using the definition
(\ref{Unm}) and the hard-wall model of side boundaries, to
ascertain the equality

\begin{equation}
\left<U_{nm}(x)V_m(x')\right>=0  \ .
\label{inter-intra}
\end{equation}
This allows us to couple the inter-mode potentials in Eq.~(\ref{krn_<T>})
only with each other, not affecting the trial functions $G_m^{(V)}(x,x_1)$.
The averaging procedure thus makes the operator $\langle {\hat{\cal
T}}_n\rangle$ effectively local.

Since equation (\ref{1Deq}) is one-dimensional in the space
variable, the exact Green function $G_{nn}(x,x')$ can be
represented in a form similar to that used for the trial function
$G_n^{(V)}(x,x')$, Eq.~(\ref{Green-Cochi}). Specifically, under
weak scattering conditions the function $G_{nn}(x,x')$, being
considered as a function of the first argument ($x$) only, is
composed of two slightly modulated exponential summands of the
appearance $\phi_{\pm}(x)=t_{\pm}(x)\exp(\pm ik_nx)$. By applying
the operator $\langle {\hat{\cal T}}_n\rangle$ to the functions
$\phi_{\pm}(x)$ one can factor the smooth amplitudes $t_{\pm}(x)$
out of the integral thus arriving at the result

\begin{equation}
\langle {\hat{\cal T}}_n\rangle \phi_{\pm}(x)=
\Sigma(k_n)\phi_{\pm}(x) \ ,
\label{<T>->Sigma}
\end{equation}
where

\begin{equation}
\Sigma(k_n)=\frac{\cal Q}{D}\int_L dx_1 W(x-x_1)
\exp\left[\mp ik_n(x-x_1)\right]\sum_{m=1\atop(m\neq n)}^{\infty}
\left<G_m^{(V)}(x,x_1)\right> \ .
\label{Sigma(k_n)}
\end{equation}
For deriving equation (\ref{Sigma(k_n)}), the correlation equality
was used

\begin{equation}
\left<U_{nm}(x)U_{kn}(x_1)\right>=\frac{\cal Q}{D}W(x-x_1)\delta_{mk}
\label{<UnmUmn>}
\end{equation}
which results from definition (\ref{Unm}) and the correlation
model (\ref{VrVr_2}). The sharp function $W(x-x_1)$ present in
Eq.~(\ref{Sigma(k_n)}) allows us to replace the trial functions
$G_m^{(V)}(x,x_1)$ by the unperturbed free mode Green functions
$G_m^{(0)}(x,x_1)$. Then, in the case of even $W(x)$, the factor
$\Sigma(k_n)$ acquires the simple form

\begin{equation}
\Sigma(k_n)=\frac{{\cal Q}}{D}\sum_{m=1\atop (m\neq n)}^{\infty}
\int\limits_{-\infty}^{\infty}\frac{dq}{2\pi}\widetilde{W}(q+k_n)
\widetilde{G}_m^{(0)}(q) \ .
\label{Sigma-int_q}
\end{equation}
Here $\widetilde{G}_m^{(0)}(q)$ is the Fourier transform of the
function $G_m^{(0)}(x)$, and it is independent of the sign of the
momentum $k_n$.

From the above analysis it follows that the action of the operator $\langle
{\hat{\cal T}}_n\rangle$ on the function $G_{nn}(x,x')$ is reduced to
multiplying by, in general, the complex-valued quantity $\Sigma(k_n)$.
Using the explicit form of $\widetilde{G}_m^{(0)}(q)$,

\begin{equation}
\widetilde{G}_m^{(0)}(q)=\frac{1}{(k_m+i0)^2-q^2} \ ,
\label{G(q)}
\end{equation}
for both real and imaginary parts of the mode `self-energy'
$\Sigma(k_n)=\Delta k_n^2-i/\tau_n^{(\varphi)}$ the expressions are
obtained as follows:

\begin{mathletters}
\label{T-renorm}
\begin{eqnarray}
\Delta k_n^2&=&\frac{{\cal Q}}{D}\sum_{m=1\atop (m\neq n)}^{\infty}
{\cal P}\!\!\!\int_{-\infty}^{\infty}\frac{dq}{2\pi}
\frac{\widetilde{W}(q+k_n)}{k_m^2-q^2} \ ,
\label{Corr_k2}
\\
\frac{1}{\tau_n^{(\varphi)}}&=&\frac{{\cal Q}}{4D}
\sum_{m=1\atop(m\neq n)}^{Nc}\frac{1}{k_m}
\left[\widetilde{W}(k_n-k_m)+\widetilde{W}(k_n+k_m)\right] \ .
\label{Atten_n}
\end{eqnarray}
\end{mathletters}
The symbol ${\cal P}$ in (\ref{Corr_k2}) denotes the principal
value. Under the conditions of weak scattering the real part
$\Delta k_n^2$ is always small, $|\Delta k_n^2|\ll k_n^2$, so it
can be disregarded without serious consequences. At the same time,
`dissipative' term (\ref{Atten_n}) plays a crucial role for the
further analysis and cannot be omitted. As a result, equation
(\ref{1Deq}) takes the form

\begin{equation}
\left[\frac{\partial^2}{\partial x^2}+\kappa_n^2
+i0-V_n(x)-\Delta{\hat{\cal T}}_n\right]
G_{nn}(x,x')=\delta(x-x') \ ,
\label{1Deq_fin}
\end{equation}
where $\kappa_n^2=k_n^2+i/\tau_n^{(\varphi)}$ and $\Delta{\hat{\cal
T}}_n={\hat{\cal T}}_n-\langle{\hat{\cal T}}_n\rangle$.  Thus, for the
analysis of equation (\ref{1Deq_fin}) we shall regard the set of
renormalized energies $\kappa_n^2$ ($n=1,2,\ldots$) as representing the new
`unperturbed' spectrum of the system, instead of the original spectrum
\{$k_n^2$\}. The perturbation theory can now be developed making use of the
appropriate zero-mean potentials $V_n(x)$ and $\Delta{\hat{\cal T}}_n$.

Note the difference in summation regions for Eqs.~(\ref{Corr_k2})
and (\ref{Atten_n}). The summation in (\ref{Atten_n}) turns out to
be restricted by the number of {\em conducting channels} (extended
waveguide modes), because only for $n\leq N_c$ the
disorder-averaged trial Green functions in Eq.~(\ref{Sigma(k_n)})
are essentially complex-valued (see Eq.~(\ref{moments}) at
$\mu=1$). The level broadening $1/\tau_n^{(\varphi)}$ implies
obligatory presence in the conductor of other extended modes
besides the $n$th mode itself. In the case of an extremely narrow
strip with $N_c=1$ the sum (\ref{Atten_n}) contains no terms, and
thus the system should exhibit true one-dimensional properties.
Specifically, the electrons in such systems can be transferred
within two regimes only, {\em ballistic} and {\em localized}, and
the conductance should go down {\em exponentially} with the length
$L$ exceeding the localization length $\xi_1$.\cite{MakTar98}

On increasing the conductor width, as soon as the wire ceases to
be single-mode ($N_c\geq 2$), the situation changes drastically.
The $n$th-mode spectrum acquires the level broadening
(\ref{Atten_n}) and is subjected to both the potentials $V_n(x)$
and $\Delta{\hat{\cal T}}_n$. The physical reason for spectrum
`complexification' with the availability of more than one extended
mode in the conductor is actually the randomization of the {\em
spatial} phase of the electron wavefunction in going between the
states with different mode energies and, consequently, mode
momenta. It is just the uncertainty of those momenta that destroys
the spatial coherence of one-dimensional quantum waves governed by
equation (\ref{1Deq_fin}). This destruction prevents
interferential localization of the true mode states of conducting
electrons contrary to their trial states.

To evaluate the `phase-breaking' effect of the term
(\ref{Atten_n}) note that at any $N_c>1$ the estimate
$1/\tau_n^{(\varphi)}\sim{\cal Q}$ is valid. In particular, in the
extreme case of a multimode conductor ($N_c\gg 1$) with point-like
scatterers, replacing the summation in Eq.~(\ref{Atten_n}) by an
integration we obtain

\begin{equation}
1/\tau_n^{(\varphi)}\approx{\cal Q}/4 \ .
\label{Ph-Br_Nc>>1}
\end{equation}
It is noteworthy that for a large number of conducting sub-bands
each level width becomes independent of the mode number $n$ and
can thus be thought of as a universal dephasing rate inherent to
the 2D conductor in general.  Although it is commonly believed
that the phase breaking stems from inelastic scattering, the level
broadening (\ref{Atten_n}) has nothing to do with such a
scattering mechanism. The only important reason for the existence
of an imaginary part of the mode energy is that the conductor has
to possess more than one extended mode.

Besides the phase breaking contribution $1/\tau_n^{(\varphi)}$, the role of
varying potentials $V_n(x)$ and $\Delta{\hat{\cal T}}_n$ should also be
studied.  These zero-mean potentials can be assessed through evaluation of
the corresponding Born scattering rates $1/\tau_n^{(V)}$ and
$1/\tau_n^{(\cal T)}$. Estimation of the operator norms $\|\hat V_n\|^2$
and $\|\Delta{\hat{\cal T}}_n\|^2$ yields

\begin{mathletters}
\label{Born_rates}
\begin{eqnarray}
\frac{\tau_n^{(V)}}{\tau_n^{(\cal T)}}&\sim&
N_c\text{min}\left(1,\frac{L}{N_c\ell}\right) \ ,
\label{estim_V/T} \\
\frac{\tau_n^{(\varphi)}}{\tau_n^{(\cal T)}}&\sim&
\frac{1}{\cos^2\vartheta_n}\text{min}\left(1,\frac{L}{N_c\ell}\right) \ .
\label{estim_phi/T}
\end{eqnarray}
\end{mathletters}
It hence follows that scattering caused by the non-local potential
$\Delta{\hat{\cal T}}_n$ is more efficient than that attributed to the
local intra-mode potential $V_n(x)$.

On the other hand, the inter-mode scattering caused by the
operator potential ${\hat{\cal T}}_n$ in Eq.~(\ref{1Deq}) is
already taken largely into account through the imaginary
renormalization of the mode energy in (\ref{1Deq_fin}). From
(\ref{estim_phi/T}) it can be seen that the scattering rate
$1/\tau_n^{(\cal T)}$ is always small compared to the level width
(\ref{Atten_n}) provided that the wire is not extremely long in
the $x$-direction, i.e. if the length $L$ does not fall into the
interval $L\gg N_c\ell$. Yet even within this interval the
quantity $1/\tau_n^{(\cal T)}$ cannot exceed the level broadening.
If that is the case, the search for strong Anderson localization
at any length of the multimode ($N_c\geq 2$) conducting strip has
no sense even without any inelastic scattering mechanisms.
Probably, it is necessary to invent some extra conditions to make
the interferential localization possible in such conductors, e.g.,
magnetic field,\cite{Malin97} surface roughness,\cite{nieto98}
etc.

\section{Calculation of the conductance}
\label{Eval_Cond}

Equations (\ref{1Deq_fin}) and (\ref{trial_Gn}) provide a way to
perform a direct calculation of the disorder-averaged conductance
(\ref{G(L)_AV}). In this section, to avoid cumbersome expressions
the results will be given for the case of a large number of
conducting channels, $N_c\gg 1$. Nevertheless, all the estimates,
as well as the final formulae, are valid for an arbitrary finite
number of modes, $N_c\gtrsim 1$. In what follows, two summands in
equation (\ref{G(L)_AV}) will be considered separately inasmuch as
the calculation techniques and the contribution of these terms in
the total conductance differ significantly. The first summand,
$\left<g^{(1)}(L)\right>$, will be conventionally referred to as
the diagonal conductance, since it does not clearly contain any
quantities characteristic of the inter-mode scattering. The second
term in (\ref{G(L)_AV}), $\left<g^{(2)}(L)\right>$, will be
conventionally referred to as the non-diagonal conductance.

Note that the effect of phase breaking, which manifests itself
strongly through complexification of the excitation spectrum at
$N_c>1$, enables one to obtain the solution of equation
(\ref{1Deq_fin}) perturbatively in the potentials $V_n$ and
$\Delta{\hat{\cal T}}_n$, i.e. to neglect, in the leading
approximation, the interference of quantum waves multiply
scattered by those potentials. It follows from estimates
(\ref{Born_rates}) that such an approach is absolutely justified
for conductors of length $L\ll N_c\ell$. Yet even in the case of a
Q1D wire of length $L\gg N_c\ell$, when the potential
$\Delta{\hat{\cal T}}_n$ in Eq.~(\ref{1Deq_fin}) cannot be
disregarded as follows from estimation (\ref{estim_phi/T}), the
effect of this potential will be shown to be negligibly small.

\subsection{The diagonal conductance}
\label{diag_cond}

The presence of the local potential $V_n(x)$ in
Eq.~(\ref{1Deq_fin}) does not substantially complicate the
calculation of the conductance. By applying the method of
Ref.~\onlinecite{MakTar98}, this potential can be taken into
account with just the same accuracy as was done diagrammatically
in Ref.~\onlinecite{berezinski73}. Accounting for this local
potential leads to purely interferential corrections, which we are
not concerned with in this paper.

As regards the operator potential $\Delta{\hat{\cal T}}_n$, to
treat it perturbatively keeping in mind the estimate
(\ref{Born_rates}) it is advantageous to go over from the
differential equation (\ref{1Deq_fin}) to the consequent integral
one,

\begin{equation}
G_{nn}(x,x')=G_{nn}^{(0)}(x,x')+\Big(\hat G_{nn}^{(0)}
\Delta{\hat{\cal T}}_n\hat G_{nn}\Big)(x,x') \ ,
\label{Gnn_Dyson}
\end{equation}
Here $G_{nn}^{(0)}(x,x')$ obeys equation (\ref{1Deq_fin}) with
$\Delta{\hat{\cal T}}_n=0$. In the leading approximation in the
parameter $L/N_c\ell\ll 1$, the function $G_{nn}^{(0)}$ has the simple
`unperturbed' form

\begin{equation}
G_{nn}^{(0)}(x,x')=\frac{1}{2ik_n}
\exp\Big\{\big[ik_n-1/l_n^{(\varphi)}\big]|x-x'|\Big\}
\label{Gnn^0}
\end{equation}
which nonetheless includes most of the inter-mode-scattering
effects. In Eq.~(\ref{Gnn^0}) the mode coherence length
$l_n^{(\varphi)}$ is associated with the $n$th level broadening
(\ref{Atten_n}), namely $l_n^{(\varphi)}=2k_n\tau_n^{(\varphi)}$.
In the limit $N_c\gg 1$ its value equals

\begin{equation}
l_n^{(\varphi)}=4\ell\cos\vartheta_n \ .
\label{l_n-approx}
\end{equation}
Substitution of the Green function (\ref{Gnn^0}) into the first
summand of Eq.~(\ref{G(L)_AV}) readily gives the diagonal part of
the conductance

\begin{equation}
\big<g^{(1)}(L)\big>=\sum_{n=1}^{N_c}\frac{l_n^{(\varphi)}}{L}
\left[1-\frac{l_n^{(\varphi)}}{L}\exp\left(-\frac{L}{l_n^{(\varphi)}}\right)
\sinh\frac{L}{l_n^{(\varphi)}}\right] \ .
\label{g(1)}
\end{equation}
In the limit $N_c\gg 1$, by replacing the sum in Eq.~(\ref{g(1)})
with the integral and substituting the coherence length in the
form (\ref{l_n-approx}), we arrive at the following asymptotic
expressions for the conductance (\ref{g(1)}):

\begin{mathletters}
\label{g(1)_asymp}
\begin{eqnarray}
L\ll\ell&\qquad\text{---}\qquad&
\big<g^{(1)}(L)\big>\approx
N_c\left(1-\frac{\pi}{12}\frac{L}{\ell}\right) \ ,
\label{g(1)_ball} \\
L\gg\ell&\qquad\text{---}\qquad&
\big<g^{(1)}(L)\big>\approx \pi N_c\ell/L \ .
\label{g(1)_diff}
\end{eqnarray}
\end{mathletters}

In the limit of $L\gg N_c\ell$, it follows from
(\ref{estim_phi/T}) that the zero-order approximation in
$\Delta{\hat{\cal T}}_n$ is, strictly speaking, insufficient for
calculating the function $G_{nn}(x,x')$. Nevertheless, since the
scattering rate $1/\tau_n^{(\cal T)}$ can only be less than or of
the order of the level width $1/\tau_n^{(\varphi)}$, the
appropriate correction to the conductance can be reasonably
estimated by substituting into $g^{(1)}(L)$ the function
$G_{nn}(x,x')$ obtained from (\ref{Gnn_Dyson}) to the first order
in $\Delta{\hat{\cal T}}_n$. A simple but tedious calculation
brings about the following estimation of the corresponding
correction to the conductance:

\begin{equation}
\big<\Delta g^{(1)}(L)\big>\sim \left(\frac{\ell}{L}\right)^2 \ .
\label{Delta_g(1)}
\end{equation}
This quantity is apparently small compared with (\ref{g(1)_diff}).
The result in (\ref{Delta_g(1)}) indicates that one may account
for the inter-mode scattering in the problem under consideration
by means of the phase breaking factor which shows itself only in
smearing of the mode energy levels.

\subsection{The non-diagonal term of the conductance}
\label{nondiag_cond}

When calculating the second term $\big<g^{(2)}(L)\big>$ in
Eq.~(\ref{G(L)_AV}) two essentially different correlators have to be
evaluated, the first containing the trial mode Green functions and the
second composed of the exact ones.  As regards the correlator of the trial
functions $G_m^{(V)}$, from result (\ref{moments}) it is clear that at
$L\ll N_c\ell$ those functions can be replaced by the free ones,
independently of the mode. Using the function $G_{nn}$ in the form
(\ref{Gnn^0}) we obtain

\begin{equation}
\big<g^{(2)}(L)\big>=-\frac{{\cal Q}}{4L^2D}
\sum_{n=1}^{N_c}\frac{\left(l_n^{(\varphi)}\right)^3}{k_n}
\exp\left(-\frac{L}{l_n^{(\varphi)}}\right)
\left[\frac{L}{l_n^{(\varphi)}}\cosh\frac{L}{l_n^{(\varphi)}}-
\sinh\frac{L}{l_n^{(\varphi)}}\right]
\sum_{m=1\atop(m\neq n)}^{N_c}\frac{1}{k_m} \ .
\label{g(2)}
\end{equation}
At $N_c\gg 1$ this expression is substantially simplified giving the
asymptotics as follows:

\begin{mathletters}
\label{g(2)_assymp}
\begin{eqnarray}
L\ll\ell \hspace{.5cm} &\qquad\text{---}\qquad&
\big<g^{(2)}(L)\big>\approx -\frac{\pi}{24}N_c\frac{L}{\ell} \ ,
\label{g(2)_ball} \\
\ell\ll L\ll N_c\ell &\qquad\text{---}\qquad&
\big<g^{(2)}(L)\big>\approx -\pi N_c\ell/2L  \ .
\label{g(2)_diff}
\end{eqnarray}
\end{mathletters}

In the case of extremely long conductors with $L\gg N_c\ell$, the trial
Green functions in (\ref{G(L)_AV}) cannot be replaced by unperturbed ones
so that the effect of 1D localization should be taken into account
properly. The `density-current' correlator standing in (\ref{G(L)_AV}) was
studied in Ref.~\onlinecite{FreylTaras91}. The result obtained there
closely corresponds to analogous results for the `density-density' and
`current-current' correlators. The main feature of the correlator is that
it decays exponentially at the localization length
$\xi_m=4L_b^{(V)}(m)$. Keeping this in mind, it is not difficult to
estimate the non-diagonal term in (\ref{G(L)_AV}) as

\begin{equation}
\big<g^{(2)}(L)\big>\sim
\left(\frac{N_c\ell}{L}\right)^2 \ .
\label{g(2)_loc}
\end{equation}

\section{Results and discussion}
\label{DISCUSSION}

While comparing results (\ref{g(2)_assymp}) and (\ref{g(2)_loc})
with those given by equations (\ref{g(1)_asymp}) it can be seen
that the non-diagonal part of the conductance is not small
relative to the diagonal one only if the conductor length falls
within the interval $\ell\ll L\ll N_c\ell$. In this case the
non-diagonal term is half the size of its diagonal counterpart and
has the opposite sign. Hence in the whole range of the conductor
length (with the width being kept constant) the conductance is
described by the following asymptotic expressions:

\begin{equation}
\begin{array}{rll}
\text{(i)} & L<\ell\ : & \big<g(L)\big>\approx N_c \\[6pt]
\text{(ii)} & \ell\ll L\ll N_c\ell\ :\qquad & \big<g(L)\big>\approx
(\pi/2) N_c\ell/L\gg 1 \\[6pt]
\text{(iii)} & N_c\ell\ll L\ : & \big<g(L)\big>\approx
\pi N_c\ell/L\ll 1 \ .
\end{array}
\label{g(L)_fin}
\end{equation}
Strictly speaking, result (\ref{g(L)_fin}) is valid exactly when
the number of channels is large, $N_c\gg 1$. Nevertheless, even in
the case of $N_c\gtrsim 1$ only an insignificant difference
occurs, produced by the dependence of the coherence length
$l_n^{(\varphi)}$ on the mode number $n$. This dependence differs
somewhat in the non-semi-classical limit from that given by
equation (\ref{l_n-approx}), but the difference reduces to a
numerical factor of the order of unity only.

The result given by Eq.~(\ref{g(L)_fin}) allows one to distinguish
three regimes of charge transport in a quantum conductor,
depending on its aspect ratio. Regime (i) corresponds to entirely
ballistic transport, both from the semi-classical and quantum
points of view. The result obtained exhibits a natural stepwise
dependence of the conductance on the transverse size of the wire.

In regimes (ii) and (iii) the semi-classical motion should be
regarded as diffusive. This opinion is consistent with the
conventional view of classical diffusion since the mean free path
$\ell$ is small compared with the sample length $L$, although it
can be in arbitrary relation to the conductor width.  Furthermore,
such an interpretation is supported by the ohmic, i.e. inversely
proportional to $L$, dependence of the conductance in both of the
indicated regimes. At the same time, it should be particularly
emphasized that only in regime (ii), commonly called the {\em weak
localization} regime, is the result given by the classical kinetic
theory reproduced exactly.  The expression for the diffusion
coefficient in regime (iii) differs from that pertinent to regime
(ii) by a factor of two.

Regime (iii) is often called {\em localized}, because it is
usually supposed that in wires of such a length (commonly referred
to as Q1D systems) the Anderson localization should manifest
itself to a considerable extent, thus leading to an exponential
fall in the conductance.  However, from the above analysis it
follows that conductors with more than one quantum channel
interconnected through scattering mechanisms, even elastic ones,
should not exhibit an exponential dependence of kinetic
coefficients on the sample length. Such a behaviour is
characteristic for the single-mode wires only, which is entirely
consistent with theoretical predictions for one-dimensional
disordered systems.

To associate findings of this paper with convictions that have
prevailed hitherto, it is helpful to examine the electron
transport in regimes (ii) and (iii) starting with the trial
waveguide states governed by the homogeneous equation
(\ref{trial_Gn}). Those states are certainly fictitious, they
would exist provided the inter-mode scattering were disregarded.
If so, the system would indeed represent a set of $N_c$
independent one-dimensional conducting channels where the true
interferential localization should take place as a result of
direct intra-mode scattering from the potentials $V_n$. For all of
the channels, a hierarchy of localization lengths would exist
similar to that representative of equation (\ref{Lfb_estim}). The
length region in (ii) corresponds to the condition when the
majority of the trial states are extended. In contrast, in regime
(iii) all of those states are localized, which is consistent with
the expectation of an exponential fall of the conductance with the
growth of the sample length.

In reality one certainly cannot disregard the inter-mode
scattering in the case of arbitrary quenched disorder. That
scattering results in quite strong coupling of the channels or,
rather, the trial mode states. The coupling has shown itself
through the complexification of true mode spectrum in
Eq.~(\ref{1Deq_fin}). Such a complexification suggests that for
any extended mode in a multi-mode strip all other extended modes
can be thought of as a phase-breaking reservoir destroying quantum
interference and hence strong (exponential) localization. Only
weak localization corrections due to the {\em local} intra-mode
potentials can be detected in the both of the diffusive regimes
(\ref{g(L)_fin}). A comprehensive analysis of the matter is beyond
the scope of this paper and will be presented elsewhere.

The existence of different diffusion regimes (ii) and (iii) can be
interpreted as the dependence of the diffusion coefficient on the
conductor aspect ratio. This dependence can hardly be extracted in
the framework of the semi-classical approach. It results from the
fact that under a gradual transition from regime (ii) to (iii),
with the growth of the conductor length, the {\em trial waveguide
states} undergo sequential localization. This should reduce the
probability for the corresponding excitations to leave the
conductor through the current terminals, whereas the probability
of their scattering into other extended modes should increase.
When all the trial states become finally localized, the diffusion
coefficient stabilizes at the value corresponding to regime (iii).
The dimensionless conductance of such a long wire is less than
unity as a consequence of the conventional Ohm's law. It seems
that this smallness has previously been the reason to suppose all
genuine states in Q1D conductors to be localized.

\acknowledgments

The author is grateful to N. M. Makarov for stimulating
discussions, A. V. Moroz for help in the interpretation of the
results and K. Ilyenko for reading the manuscript.

\appendix

\section{On regularity of the operator $\protect\hat{\sf K}$,
E\protect\lowercase{q}.~(\protect\ref{HAT_K})}
\label{K_exist}

To ascertain that unity is not among the characteristic numbers of
the operator $\hat{\sf R}$, let us take advantage of the operator
identity

\begin{equation}
\log\det(\openone-\hat{\sf R})=\text{Tr}\log(\openone-\hat{\sf R})=
\text{Tr}\sum_{k=1}^{\infty}\frac{(-1)^k}{k}\hat{\sf R}^k \ .
\label{Simil}
\end{equation}
The last equality in (\ref{Simil}) presumes that the operator norm is
limited by $\|\hat{\sf R}\|<1$. This is entirely consistent with the
estimate (\ref{normR}).

Consider the traces of the first two terms in sum (\ref{Simil}),

\begin{eqnarray}
\text{Tr}\,\hat{\sf R}&=&\sum_{n=1}^{\infty}
\int_Ldx\,G_n^{(V)}(x,x)U_{nn}(x) \ ,
\label{First}
\\
\text{Tr}\,{\hat{\sf R}}^2&=&\sum_{n,m=1}^{\infty}
\int\!\!\!\!\int_Ldxdx'\,G_n^{(V)}(x,x')U_{nm}(x')
G_m^{(V)}(x',x)U_{mn}(x) \ .
\label{Second}
\end{eqnarray}
Since at $n>N_c$ the function $G_n^{(V)}(x,x')$ in the case of
weak scattering has the form (\ref{G_evan}), it is easy to
conclude that the divergence of the logarithm in (\ref{Simil}) can
arise from the first term, given in (\ref{First}), provided
$U_{nn}(x)\not\equiv 0$. The divergence stems from locality, i.e.
coincidence of the arguments, of the Green functions. It manifests
itself not only on average but also at a given realization of the
random potential entering this term. By separating the diagonal,
i.e.  intra-mode, potential $V_n(x)\equiv U_{nn}(x)$ and making
the matrix $\|U_{nm}\|$ off-diagonal we prevent the singularity of
the operator $\hat{\sf K}$ given by Eq.~(\ref{HAT_K}).

\section{Statistical moments of the trial Green functions}
\label{MOMENTS-GN}

To perform the ensemble-averaging of the random function
$\Phi_{\mu}(x,x'|n)=\big[G_n^{(V)}(x,x')\big]^\mu$,
$\mu\in\aleph$, in accordance with representation
(\ref{Green-Cochi}) and (\ref{psi-pm}), we first decompose the
function $G_n^{(V)}(x,x')$ into the sum of four terms each
containing narrow packets only of spatial harmonics with phases
close to $\pm k_n(x\pm x')$.  In doing so one should use the
asymptotic expression for the Wronskian $\bbox{\cal W}_n$,

\begin{equation}
\bbox{\cal W}_n=2ik_n\big[\pi_{+}(x|n)\pi_{-}(x|n)+
\gamma_{+}(x|n)\gamma_{-}(x|n)\big] \ .
\label{W_appr}
\end{equation}
This results from the assumption of smoothness of the amplitude
functions in Eq.~(\ref{psi-pm}). Then, after substituting
(\ref{psi-pm}) and (\ref{W_appr}) into (\ref{Green-Cochi}), the
function $G_n^{(V)}(x,x')$ can be represented in the form of a
scalar product

\begin{equation}
G_n^{(V)}(x,x')=\pmatrix{{\rm e}^{ik_nx}&\!\!\!\!;\ {\rm e}^{-ik_nx}}
{{\widetilde G_1\ \widetilde G_3 }\choose
{\widetilde G_4\ \widetilde G_2}}
{{{\rm e}^{-ik_nx'}} \choose {{\rm e}^{ik_nx'}} } \ .
\label{scal_prod}
\end{equation}
Here $\widetilde G_i(x,x'|n)$ are the smooth amplitudes constructed from
the envelopes $\pi_{\pm}(x|n)$ and $\gamma_{\pm}(x|n)$ as follows:

\begin{mathletters}
\label{G_matr}
\begin{eqnarray}
\widetilde G_1(x,x'|n)&=&\frac{-i}{2k_n}A_n(x)
\left[\Theta_+\frac{\pi_{-}(x'|n)}{\pi_{-}(x|n)}-
\Theta_-\frac{\gamma_{+}(x'|n)}{\pi_{+}(x|n)}\Gamma_{-}(x|n)\right] \ ,
\label{G_1} \\[6pt]
\widetilde G_2(x,x'|n)&=&\frac{-i}{2k_n}A_n(x)
\left[\Theta_-\frac{\pi_{+}(x'|n)}{\pi_{+}(x|n)}
-\Theta_+\Gamma_{+}(x|n)\frac{\gamma_{-}(x'|n)}{\pi_{-}(x|n)}\right] \ ,
\label{G_2} \\[6pt]
\widetilde G_3(x,x'|n)&=&\frac{-1}{2k_n}A_n(x)
\left[\Theta_+\frac{\gamma_{-}(x'|n)}{\pi_{-}(x|n)}+
\Theta_-\frac{\pi_{+}(x'|n)}{\pi_{+}(x|n)}\Gamma_{-}(x|n)\right] \ ,
\label{G_3} \\[6pt]
\widetilde G_4(x,x'|n)&=&\frac{-1}{2k_n}A_n(x)
\left[\Theta_-\frac{\gamma_{+}(x'|n)}{\pi_{+}(x|n)}
+\Theta_+\Gamma_{+}(x|n)\frac{\pi_{-}(x'|n)}{\pi_{-}(x|n)}\right] \ .
\label{G_4}
\end{eqnarray}
\end{mathletters}
The notation used in Eqs.~(\ref{G_matr}) is
$$ \hspace{-1cm}
A_n(x)=\left[1+\Gamma_{+}(x|n)\Gamma_{-}(x|n)\right]^{-1} \ ,
\qquad  \Gamma_{\pm}(x|n)=\frac{\gamma_{\pm}(x|n)}{\pi_{\pm}(x|n)}
\ , \qquad \Theta_{\pm}=\Theta[\pm(x-x')] \ .
$$

Before averaging the functions (\ref{G_matr}) over the random
potential, let us note some useful features of the dynamic system
(\ref{PiGamma_eq}). Since $\pi_{\pm}(x|n)$ and $\gamma_{\pm}(x|n)$
are the {\em causal} functionals of the fields $\eta_n(x)$,
$\zeta_{n\pm}(x)$ and $\zeta_{n\pm}^*(x)$, they are determined by
the values of those fields on the intervals $(x,L/2]$ and
$[-L/2,x)$ for the functionals labelled by the indexes ($+$) and
($-$), correspondingly. The Green function elements
(\ref{G_matr}), which will be subjected to ensemble averaging, are
constructed in such a fashion that supports of the random
functions entering the functionals of `plus' and `minus' type do
not meet. Due to the random fields being effectively
$\delta$-correlated, see Eqs.~(\ref{Eta_Zeta}) and (\ref{F_l}),
averaging of the functionals with different sign indexes can be
performed independently.

It also follows from equations (\ref{PiGamma_eq}) and conditions
(\ref{In_cond}) that all terms of functional series for
$\pi_{\pm}(x|n)$ contain an equal number of fields $\zeta_{n\pm}$
and $\zeta_{n\pm}^*$, whereas $\gamma_{\pm}(x|n)$ has an extra
functional factor $\zeta_{n\pm}$. Since for weak scattering all
fields $\eta_n(x)$, $\zeta_{n\pm}(x)$ and $\zeta_{n\pm}^*(x)$ are
approximately {\em Gaussian} random processes, only the first
summands in square brackets of Eqs.~(\ref{G_1}) and (\ref{G_2})
remain non-zero after averaging, while the quantities
$\big<\widetilde G_{3,4}\big>$ vanish. By the same arguments, the
factor $A_n(x)$ in (\ref{G_matr}) can be replaced by unity.

In view of the statistical independence of the functions
$\eta_n(x)$ and $\zeta_{n\pm}(x)$, it is convenient to average
over the real field $\eta_n(x)$ already at the initial stage. To
that end, it is advantageous to perform the following phase
transformation of the amplitudes $\pi_{\pm}$ and $\gamma_{\pm}$:

\begin{eqnarray}
\pi_{\pm}(x|n)&=&\widetilde\pi_{\pm}(x|n)\exp\bigg[\pm i\int_x^{\pm L/2}
\!\!\!\!\eta_n(x_1)\,dx_1\bigg] \ , \nonumber
\\[-7pt]\label{phase_ren}\\[-7pt]
\gamma_{\pm}(x|n)&=&\widetilde\gamma_{\pm}(x|n)\exp\bigg[\mp i\int_x^{\pm
L/2} \!\!\!\!\eta_n(x_1)\,dx_1\bigg] \ .
\nonumber
\end{eqnarray}
The new amplitudes $\widetilde\pi_{\pm}$ and $\widetilde\gamma_{\pm}$ obey
the equations

\begin{equation}
\begin{array}{ccl}
&&\widetilde\pi'_{\pm}(x|n)\pm\widetilde\zeta_{n\pm}^*(x)
\widetilde\gamma_{\pm}(x|n) =0\ , \\[6pt]
&&\widetilde\gamma'_{\pm}(x|n)\pm\widetilde\zeta_{n\pm}(x)
\widetilde\pi_{\pm}(x|n) =0\ ,
\end{array}
\label{tildePiGamma_eq}
\end{equation}
where the random field $\widetilde\zeta_{n\pm}(x)$ is related to
$\zeta_{n\pm}(x)$ by the equality

\begin{equation}
\widetilde\zeta_{n\pm}(x)=\zeta_{n\pm}(x)\exp\bigg[\pm 2i\int_x^{\pm L/2}
\!\!\!\!\eta_n(x_1)\,dx_1\bigg] \ .
\label{tilde-zeta}
\end{equation}
This latter condition does not modify correlation properties of
the backscattering fields, Eqs.~(\ref{Eta_Zeta}). Then performing
a Fourier transformation of the function $\Phi_{\mu}(x,x'|n)$ over
$x'$, we arrive at the expression conveniently decomposed into the
sum of `plus' and `minus' functionals,

\begin{equation}
\widetilde\Phi_{\mu}(x,q|n)=\left(-
\frac{1}{2k_n}\right)^\mu{\rm e}^{iqx}
\left[\widetilde\Phi_{\mu}^{(+)}(x,q|n)+
\widetilde\Phi_{\mu}^{(-)}(x,q|n)\right] \ .
\label{Phi_N+-}
\end{equation}
Here the functions $\widetilde\Phi_{\mu}^{(\pm}$ are given by

\begin{equation}
\widetilde\Phi_{\mu}^{(\pm)}(x,q|n)=\pm \int_x^{\pm L/2}
\!\!\!\!dx_1\left[\frac{\widetilde\pi_{\pm}(x_1|n)}
{\widetilde\pi_{\pm}(x|n)}\right]^\mu
\exp\left[-iq(x-x_1)+i\mu k_n|x-x_1| \pm i\mu\int_{x_1}^{x}
\eta_n(x_2)\,dx_2\right] \ .
\label{Phi^(pm)}
\end{equation}
Averaging functions (\ref{Phi^(pm)}) over the random field $\eta_n(x)$
with the use of (\ref{EtaEta}) readily yields

\begin{equation}
\widetilde\Phi_{\mu}^{(\pm)}(x,q|n)=\pm \int_x^{\pm L/2}
\!\!\!\!dx_1\left[\frac{\widetilde\pi_{\pm}(x_1|n)}
{\widetilde\pi_{\pm}(x|n)}\right]^\mu
\exp\left\{-iq(x-x_1)+\left[i\mu k_n -\frac{\mu^2}{L_f^{(V)}(n)}
\right]|x-x_1|\right\} \ .
\label{Phi^(pm)av-eta}
\end{equation}

To then perform averaging over the fields $\zeta_{n\pm}(x)$ it is
convenient to use the dynamic equations for the functions
$\widetilde\Phi_{\mu}^{(\pm)}(x,q|n)$ and
$\widetilde\Gamma_{\pm}(x|n)=
{\widetilde\gamma_{\pm}(x|n)}/{\widetilde\pi_{\pm}(x|n)}$. They
read

\begin{equation}
\mp\frac{d\widetilde\Phi_{\mu}^{(\pm)}(x,q|n)}{dx}=
1-\left[\frac{\mu^2}{2L_f^{(V)}(n)}-i\mu k_n\mp iq\right]
\widetilde\Phi_{\mu}^{(\pm)}(x,q|n)-
\mu\widetilde\zeta_{n\pm}^*(x)\widetilde\Gamma_{\pm}(x|n)
\widetilde\Phi_{\mu}^{(\pm)}(x,q|n) \ ,
\label{Phi^pm_eq}
\end{equation}
\begin{equation}
\pm \frac{d\widetilde\Gamma_{\pm}(x|n)}{dx}=
-\widetilde\zeta_{n\pm}(x)+
\widetilde\zeta_{n\pm}^*(x)\widetilde\Gamma_{\pm}^2(x|n) \ .
\label{Gamma_pm-eq}
\end{equation}
These equations stem from definitions (\ref{Phi^(pm)}) and system
(\ref{tildePiGamma_eq}) along with the obvious `initial' conditions

\begin{equation}
\widetilde\Phi_{\mu}^{(\pm)}(\pm L/2,q|n)=0 \ ,\hspace{1.5cm}
\widetilde\Gamma_{\pm}(\pm L/2|n)=0 \ .
\label{In-cond}
\end{equation}
Averaging of (\ref{Phi^pm_eq}) with the use of Furutsu-Novikov
formula\cite{Klyatskin86} gives the equation

\begin{equation}
\frac{d\big<\widetilde\Phi_{\mu}^{(\pm)}(x,q|n)\big>}{dx}=
1-\left[\frac{\mu}{2}\left(\frac{\mu}{L_f^{(V)}(n)}+
\frac{1}{L_b^{(V)}(n)}\right)-i\mu k_n\mp iq\right]
\big<\widetilde\Phi_{\mu}^{(\pm)}(x,q|n)\big> \ ,
\label{Phi_fineq}
\end{equation}
from which the result (\ref{moments}) arises immediately.



\begin{thebibliography}{25}
\bibitem{myremark_1}
Although universality of the one-parameter scaling was quite long
challenged,\cite{abrikos81,kravts_lerner84} attempts were made
(and still persist) to improve the scaling approach owing to its
relative convenience and
simplicity.\cite{shreib_ottom92,dobrosavl97}
\end{thebibliography}
\end{document}